# Impact and Relevance of Cognition Journal in the Field of Cognitive Science: An Evaluation

M Sadik Batcha[1], Younis Rashid Dar[2], Muneer Ahmad[3*],

[1]Mentor and Research Supervisor, Professor and University Librarian, Annamalai University, Annamalai nagar, India

[2]Ph.D Research Fellow, Department of Linguistics, University of Kashmir, Hazratbal Srinagar, India

[3*]Ph.D Research Scholar, Department of Library and Information Science, Annamalai University, Annamalai nagar, India

*Corresponding Author:* Muneer Ahmad, Ph.D Research Scholar, Department of Library and Information Science, Annamalai University, Annamalai nagar, India, , Email: muneerbangroo@gmail.com

**ABSTRACT**

*This study aims to present a scientometric analysis of the journal titled "Cognition" for a period of 20 years from 1999 to 2018. The present study was conducted with an aim to provide a summary of research activity in current journal and characterize its most aspects. The research coverage includes the year wise distribution of articles, authors, institutions, countries and citation analysis of the journal. The analysis showed that 2870 papers were published in journal of Cognition from 1999 to 2018. The study identified top 20 prolific authors, institutions and countries of the journal. Researchers from USA have been made the most percentage of contributions.*

## INTRODUCTION

Cognition as defined by (Neisser, 1967)[1] "all the processes by which sensory input is transformed, reduced, elaborated, stored , recovered and used" provides an overall sense of the term "*Cognition*". Cognition is the use of conscious mental processes underlying our ability to think, to talk, to remember, to learn from our experiences and to modify and control our behaviour accordingly. However, sometimes the terms "*Cognition*" and "*Thinking*" are taken as synonyms. The topics of cognition involve the study of perception, knowledge representation, attention, thought, Language, problem solving, decision-making, memory, consciousness and aspects of Intelligence. Studying about cognition is fundamental to understanding various mental processes and the subject of cognition finds its place among the basic and most important topics in a growing and broader discipline of cognitive psychology and related fields.

*Cognition* is an international journal that publishes theoretical and experimental papers on the study of the mind. It publishes many of the most important papers in cognitive science and is the premier international and interdisciplinary journal in the field. It is required reading for anyone who wishes to keep up to date in this exciting research area. It covers a wide variety of subjects concerning all the different aspects of cognition, ranging from biological and experimental studies to formal analysis.

Contributions from the fields of psychology, neuroscience, linguistics, computer science, mathematics, ethology and philosophy are published in the journal provided that the articles have some bearing on the functioning of the mind. In addition, the journal serves as a forum for discussion of social and political aspects of cognitive science (Martin, Tsakiris, & Wagemans, 2019)[2].

Scientometric is the study of measuring research quality and impact, understanding the processes of citations, scientific mapping fields, and the use of indicators in research policy and management (Mingers & Leydesdorff, 2015)[3]. The scientometric analysis shows the topics inside a search criterion, for example, the top countries evolution inside the criterion countries, or a list of specific author keywords inside the criterion author keywords. The temporal analysis allows us to find when a new phenomenon starts, and when it advances to a trending or emerging topic. The scientometric tools have developed a kind of algorithms to perform the temporal analysis and find the trending topics, such as strategic diagrams (Cobo, Pez-Herrera, Herrera-Viedma, & Herrera, 2012)[4] and Kleinberg's burst detection algorithm (Kleinberg, 2003)[5]. These kinds of analysis are performed in datasets that generally are extracted from a single bibliometric database, like Scopus or WoS, because, most of the tools cannot merge the information successfully from different databases. Also, there is not a longitudinal graph representation of the trending topics evolution provided by all the actual tools.

## REVIEW OF LITERATURE

In recent years, many researchers have conducted scientometric analysis in different fields. (Batcha & Ahmad, 2017)[6] analysed comparative analysis of Indian Journal of Information Sources and Services (IJISS) and Pakistan Journal of Library and Information Science (PJLIS) during 2011-2017 and studied various aspects like year wise distribution of papers, authorship pattern & author productivity, degree of collaboration pattern of Co-Authorship, average length of papers, average keywords, etc and found 138 (94.52%) of contributions from IJISS were made by Indian authors and similarly 94 (77.05) of contributions from PJLIS were done by Pakistani authors. Papers by Indian and Pakistani Authors with Foreign Collaboration are minimal (1.37% of articles) and (4.10% of articles) respectively.

(Batcha, Jahina, & Ahmad, 2018)[7] has examined scientometric analysis of the DESIDOC Journal and analyzed the pattern of growth of the research output published in the journal, pattern of authorship, author productivity, and, subjects covered to the papers over the period (2013-2017). It found that 227 papers were published during the period of study (2001-2012). The maximum numbers of articles were collaborative in nature. The subject concentration of the journal noted was Scientometrics. The maximum numbers of articles (65 %) have ranged their thought contents between 6 and 10 pages.

(Ahmad & Batcha, 2019)[8] analyzed research productivity in Journal of Documentation (JDoc) for a period of 30 years between 1989 and 2018. Web of Science database a service from Clarivate Analytics has been used to download citation and source data. Bibexcel and Histcite application software have been used to present the datasets. Analysis part focuses on the parameters like citation impact at local and global level, influential authors and their total output, ranking of contributing institutions and countries. In addition to this scientographical mapping of data is presented through graphs using VOS viewer software mapping technique.

(Ahmad, Batcha, Wani, Khan, & Jahina, 2017)[9] explored scientometric analysis of the Webology Journal. The paper analyses the pattern of growth of the research output published in the journal, pattern of authorship, author productivity, and subjects covered to the papers over the period (2013-2017). It was found that 62 papers were published during the period of study (2013-2017). The maximum numbers of articles were collaborative in nature. The subject concentration of the journal noted was Social Networking/Web 2.0/Library 2.0 and Scientometrics or Bibliometrics. Iranian researchers contributed the maximum number of articles (37.10%). The study applied standard formula and statistical tools to bring out the factual results.

(Ahmad & Batcha, 2019)[10] studied the scholarly communication of Bharathiar University which is one of the vibrant universities in Tamil Nadu. The study find out the impact of research produced, year-wise research output, citation impact at local and global level, prominent authors and their total output, top journals of publications, collaborating countries, most contributing departments and publication trends of the university during 2009 to 2018. The 10 years' publication data of the university indicate that a total of 3440 papers have been published from 2009 to 2018 receiving 38104 citations with h-index as 68. In addition the study used scientographical mapping of data and presented it through graphs using VOS viewer software mapping technique.

(Ahmad, Batcha, & Jahina, 2019)[11] quantitatively identified the research productivity in the area of artificial intelligence at global level over the study period of ten years (2008-2017). The study identified the trends and characteristics of growth and collaboration pattern of artificial intelligence research output. Average growth rate of artificial intelligence per year increases at the rate of 0.862. The multi-authorship pattern in the study is found high and the average number of authors per paper is 3.31. Collaborative Index is noted to be the highest range in the year 2014 with 3.50. Mean CI during the period of study is 3.24. This is also supported by the mean degree of collaboration at the percentage of 0.83 .The mean CC

**Impact and Relevance of Cognition Journal in the Field of Cognitive Science: An Evaluation**

observed is 0.4635. Lotka's Law of authorship productivity is good for application in the field of artificial intelligence literature. The distribution frequency of the authorship follows the exact Lotka's Inverse Law with the exponent á = 2. The modified form of the inverse square law, i.e., Inverse Power Law with á and C parameters as 2.84 and 0.8083 for artificial intelligence literature is applicable and appears to provide a good fit. Relative Growth Rate [Rt(P)] of an article gradually increases from -0.0002 to 1.5405, correspondingly the value of doubling time of the articles Dt(P) decreases from 1.0998 to 0.4499 (2008-2017). At the outset the study reveals the fact that the artificial intelligence literature research study is one of the emerging and blooming fields in the domain of information sciences.

### OBJECTIVES OF THE STUDY

The main objective of the study is to consider on the mapping of 2870 articles published by the *Cognition* journal during the period of 1999 – 2018 and the specific objectives are to identify and carry out the following factors:

- To examine the annual publications output of *Cognition* journal.
- To estimate publication density through mapping of top 20 authors, countries and institutions based on their number of research papers.
- Find out the top 20 prolific authors, institutions and countries.

### DATA SOURCE AND METHODOLOGY

The data for the present study were taken from the Clarivate analytics-Web of Science database in July 2019. A total of 2870 research publications was downloaded from 1999-2018. The data downloaded was enhanced with different parameters like title, authors, years, countries, and research institutions. Furthermore, the downloaded data were analyzed by using Bibexcel, Histcite, and VOS viewer software applications.

**Table1.** *Detail of the Important Points of the Data Sample During 1999 to 2018*

| S.No. | Details about Sample | Observed Values |
|---|---|---|
| 1 | Duration | 1999-2018 |
| 2 | Collection Span | 20 Years |
| 3 | Total No. of Records | 2870 |
| 4 | Total No. of Authors | 5233 |
| 5 | Frequently Used Words | 4550 |
| 6 | Document Types | 9 |
| 7 | Languages | 1 |
| 8 | Contributing Countries | 61 |
| 9 | Contributing Institutions | 1127 |
| 10 | Institutions with Sub Division | 2709 |
| 11 | Total Cited References | 145261 |
| 12 | Total Local Citation Scores | 5045 |
| 13 | Total Global Citation Scores | 121457 |
| 14 | H-Index | 159 |

### DISCUSSION AND RESULT

#### Evaluate the Annual Output of Publications

The data from Table 2 and graph 1 can be clearly seen that the numbers of research documents published from 1999 to 2018 shows a gradual increase in publication of research articles in the Journal. According to the publication output from the Table 2 the year wise distribution of research documents, 2008 has the highest number of research documents 255 (8.89%) with 746 (14.79%) of total local citation score and 16855 (13.88%) of total global citation score values and being prominent among the 20 years output and it stood in first rank position. The year 2016 has 249 (8.68%) research documents and it stood in second position with 116 (2.30%) of total local citation score and 2241 (1.85%) of total global citation score were scaled. It is followed by the year 2018 with 236 (8.22 %) of records and it stood in third rank position along with 8 (0.16%) of total local citation score and 337 (0.28%) of total global citation score measured. The year 2017 has 234 (8.15%) research documents and it stood in fourth position with 71 (1.41%) & 1220 (1.00%). The year 2015 has 217 (7.56%) research documents and it stood in 5th position with 151(2.99%) of total local citation score and 2987 (2.46%) of total global citation score were scaled. It has been observed from the data that the increase in publications in the journal does not necessarily imply an increase in the overall citation score of the research articles. Graph no.

**Impact and Relevance of Cognition Journal in the Field of Cognitive Science: An Evaluation**

1 presents the year wise publications and depicts the citation score. It clearly indicates on the fact that increase in publication rate is not directly linked to increase in citation score.

**Table2.** *Annual Distribution of Publications and Citations*

| S.No. | Year | Records | % | Rank | TLCS | % | Rank | TGCS | % | Rank |
|---|---|---|---|---|---|---|---|---|---|---|
| 1 | 1999 | 64 | 2.23 | 19 | 396 | 7.85 | 2 | 7390 | 6.08 | 5 |
| 2 | 2000 | 60 | 2.09 | 20 | 308 | 6.11 | 5 | 6671 | 5.49 | 8 |
| 3 | 2001 | 77 | 2.68 | 15 | 342 | 6.78 | 4 | 9034 | 7.44 | 2 |
| 4 | 2002 | 78 | 2.72 | 14 | 51 | 1.01 | 19 | 6849 | 5.64 | 7 |
| 5 | 2003 | 76 | 2.65 | 16 | 291 | 5.77 | 9 | 6337 | 5.22 | 11 |
| 6 | 2004 | 73 | 2.54 | 18 | 185 | 3.67 | 14 | 8313 | 6.84 | 3 |
| 7 | 2005 | 84 | 2.93 | 13 | 292 | 5.79 | 8 | 6533 | 5.38 | 9 |
| 8 | 2006 | 75 | 2.61 | 17 | 242 | 4.80 | 10 | 5800 | 4.78 | 13 |
| 9 | 2007 | 119 | 4.15 | 12 | 237 | 4.70 | 12 | 6497 | 5.35 | 10 |
| 10 | 2008 | 255 | 8.89 | 1 | 746 | 14.79 | 1 | 16855 | 13.88 | 1 |
| 11 | 2009 | 149 | 5.19 | 11 | 358 | 7.10 | 3 | 8018 | 6.60 | 4 |
| 12 | 2010 | 161 | 5.61 | 8 | 293 | 5.81 | 7 | 7385 | 6.08 | 6 |
| 13 | 2011 | 153 | 5.33 | 10 | 298 | 5.91 | 6 | 6320 | 5.20 | 12 |
| 14 | 2012 | 160 | 5.57 | 9 | 236 | 4.68 | 13 | 4784 | 3.94 | 14 |
| 15 | 2013 | 181 | 6.31 | 6 | 242 | 4.80 | 10 | 4494 | 3.70 | 15 |
| 16 | 2014 | 169 | 5.89 | 7 | 182 | 3.61 | 15 | 3392 | 2.79 | 16 |
| 17 | 2015 | 217 | 7.56 | 5 | 151 | 2.99 | 16 | 2987 | 2.46 | 17 |
| 18 | 2016 | 249 | 8.68 | 2 | 116 | 2.30 | 17 | 2241 | 1.85 | 18 |
| 19 | 2017 | 234 | 8.15 | 4 | 71 | 1.41 | 18 | 1220 | 1.00 | 19 |
| 20 | 2018 | 236 | 8.22 | 3 | 8 | 0.16 | 20 | 337 | 0.28 | 20 |
| | Total | 2870 | 100.00 | | 5045 | 100.00 | | 121457 | 100.00 | |

*\*TLCS = Total Local Citation Score, \*TGCS = Total Global Citation Score*

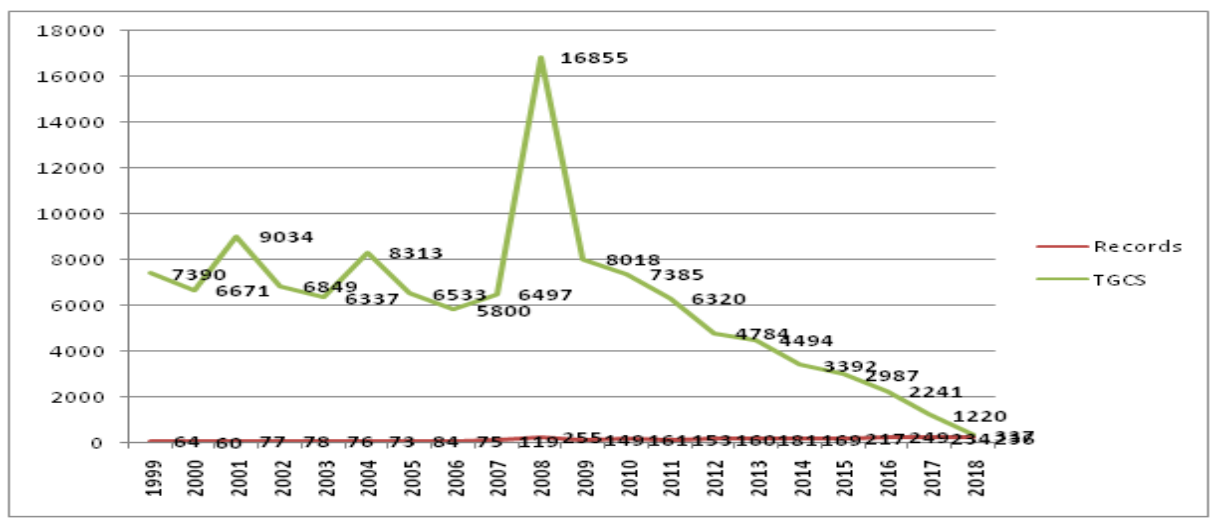

**Graph1.** *Annual Distribution of Publications and Citations*

**Analysis of the Publication Output of Top 20 Authors**

The ranking of authors of various research articles is displayed in Table 3 and figure 1. In the rank analysis the authors who have published more than 12 articles or more are considered into account to avoid a long list. It is observed that there are a total of 5233 authors for 2870 records and it shows the top 20 most productive authors during 1999-2018. Spelke ES published 25 (0.87%) articles with 2370 TGCS articles, followed by Carey S 24 (0.84%) with 1596 TGCS articles, Bloom P 19 (0.66%) with 1225 TGCS articles, Tanenhaus MK 19 (0.66%) with 1140 TGCS articles, Tomasello M 19 (0.66%) with 1496 TGCS article, Tenenbaum JB 18 (0.63%) with 737 TGCS articles, Dehaene S with 17 articles (0.59%) with 2196 TGCS and other authors have contributed comparatively less than the top seven authors during the period of study.

The data set clearly depicts that the increase in number of publications of an authors usually has a direct impact in the increase in citation score. It is found that the ranked contributors are from the following research Institutions: Harvard

**Impact and Relevance of Cognition Journal in the Field of Cognitive Science: An Evaluation**

University, University College London, MIT, Yale University, University of Illinois and so on. It could be identified from the author wise analysis, the following authors: Spelke ES, Carey S, Bloom P, Tanenhaus MK, Tomasello M and Tenenbaum JB were the most productive authors based on the number of research papers published in the Journal. The data set puts forth that the authors Spelke ES with 2370 citations, Dehaene S with 2196 citations, Aslin RN with 1779 citations and Carey S with 1596 citations.

**Table3.** *Publication output of Top 20 Authors and Citation Score*

| S.No | Author | Records | % | TLCS | TGCS |
|---|---|---|---|---|---|
| 1 | Spelke ES | 25 | 0.87 | 136 | 2370 |
| 2 | Carey S | 24 | 0.84 | 127 | 1596 |
| 3 | Bloom P | 19 | 0.66 | 80 | 1225 |
| 4 | Tanenhaus MK | 19 | 0.66 | 57 | 1140 |
| 5 | Tomasello M | 19 | 0.66 | 75 | 1496 |
| 6 | Tenenbaum JB | 18 | 0.63 | 65 | 737 |
| 7 | Dehaene S | 17 | 0.59 | 66 | 2196 |
| 8 | Haggard P | 16 | 0.56 | 35 | 883 |
| 9 | Goldin-Meadow S | 15 | 0.52 | 16 | 550 |
| 10 | Griffiths TL | 15 | 0.52 | 63 | 663 |
| 11 | Call J | 14 | 0.49 | 17 | 440 |
| 12 | Waxman SR | 14 | 0.49 | 39 | 544 |
| 13 | Baillargeon R | 13 | 0.45 | 77 | 668 |
| 14 | Pickering MJ | 13 | 0.45 | 45 | 782 |
| 15 | Aslin RN | 12 | 0.42 | 90 | 1779 |
| 16 | Fisher C | 12 | 0.42 | 44 | 561 |
| 17 | Frank MC | 12 | 0.42 | 33 | 498 |
| 18 | Gelman SA | 12 | 0.42 | 51 | 547 |
| 19 | Hauser MD | 12 | 0.42 | 35 | 894 |
| 20 | Knoblich G | 12 | 0.42 | 25 | 574 |

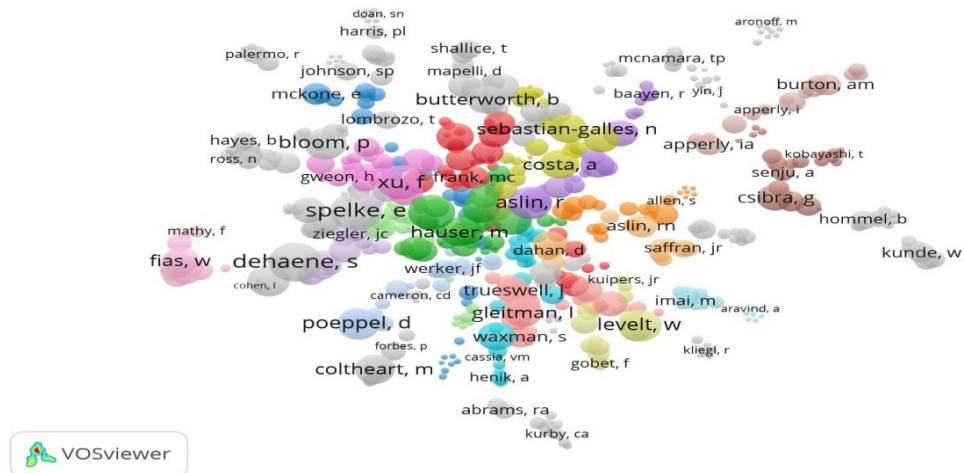

**Figure1.** *Highly Prolific Authors*

### Analysis of the Publication Output of Top 20 Institutions

The individualities of 20 most productive institutions were analyzed in this part, Institutions which published more than 41 and above publications have been considered as highly productive institutions. Table 5 summarizes articles, the global citation score, local citation score and average author per paper of the publications of these institutions. In total, 1127 institutions, including 2709 subdivisions published 2870 research papers during 1999 – 2018. The topmost twenty institutions involved in this research have published 41 and more research articles.

The mean average is 2.54 research articles per Institution. Out of 1127 institutions, top 20 institutions published 1247 (43.45%) research papers and the rest of the institution published 1623 (56.55%) research papers respectively. Based on the number of published research records the institutions are ranked.

**Impact and Relevance of Cognition Journal in the Field of Cognitive Science: An Evaluation**

Among the institution listed in Table 4 "Harvard University" holds the first rank and the institution published 149 (5.19%) research papers with 404 local and 9140 global citation scores, the average citation per paper is 61.34. The second rank holds by "University College London" the institution published 110 (3.83%) research papers with 223 local and 6613 global citation scores, the average citation per paper is 60.12. The "MIT" holds the 3rd rank, the institution published 74 (2.58%) research papers with 317 local and 4978 global citation scores, the average citation per paper is 67.27. The "Yale University" holds the 4th rank, the institution published 74 (2.58%) research papers with 197 local and 3125 global citation scores, the average citation per paper is 42.23. The "University of Illinois" holds the 5th rank, the institution published 66 (2.30%) research papers with 201 local and 2899 global citation scores, the average citation per paper is 43.92. It is clear from the analysis that following institutions: Harvard University, University College London, MIT, Yale University and University of Illinois among others were identified the most productive institutions based on the number of research papers published in the Journal. However, University of Rochester (79.70), MIT (67.27), Max Plank Institute for Psycholinguistics (65.84), Harvard University (61.34) and University College London (60.12) are the institutions with high ACPP indicating the quality work with high citation impact hence they can be recognized as the most productive institutions based on the annual citation per paper received in terms of publications.

**Table4.** *Ranking of Institutions and their Research Performance*

| S.No. | Institution | Records | % | TLCS | TGCS | ACPP |
|---|---|---|---|---|---|---|
| 1 | Harvard University | 149 | 5.19 | 404 | 9140 | 61.34 |
| 2 | University College London | 110 | 3.83 | 223 | 6613 | 60.12 |
| 3 | MIT | 74 | 2.58 | 317 | 4978 | 67.27 |
| 4 | Yale University | 74 | 2.58 | 197 | 3125 | 42.23 |
| 5 | University of Illinois | 66 | 2.30 | 201 | 2899 | 43.92 |
| 6 | Northwestern University | 64 | 2.23 | 126 | 2327 | 36.36 |
| 7 | University Chicago | 62 | 2.16 | 169 | 2979 | 48.05 |
| 8 | CNRS | 59 | 2.06 | 95 | 2493 | 42.25 |
| 9 | University Pennsylvania | 59 | 2.06 | 132 | 3055 | 51.78 |
| 10 | University of Edinburgh | 58 | 2.02 | 127 | 2250 | 38.79 |
| 11 | Stanford University | 55 | 1.92 | 149 | 3189 | 57.98 |
| 12 | University of Rochester | 53 | 1.85 | 207 | 4224 | 79.70 |
| 13 | University Calif San Diego | 52 | 1.81 | 114 | 1946 | 37.42 |
| 14 | New York University | 49 | 1.71 | 96 | 1709 | 34.88 |
| 15 | University of Oxford | 49 | 1.71 | 69 | 2039 | 41.61 |
| 16 | Max Planck Institution, Psycholinguist | 44 | 1.53 | 70 | 2897 | 65.84 |
| 17 | Radboud University Nijmegen | 44 | 1.53 | 43 | 1509 | 34.30 |
| 18 | Brown University | 43 | 1.50 | 104 | 1422 | 33.07 |
| 19 | University of Padua | 42 | 1.46 | 72 | 2420 | 57.62 |
| 20 | University of California, Berkeley | 41 | 1.43 | 121 | 1506 | 36.73 |

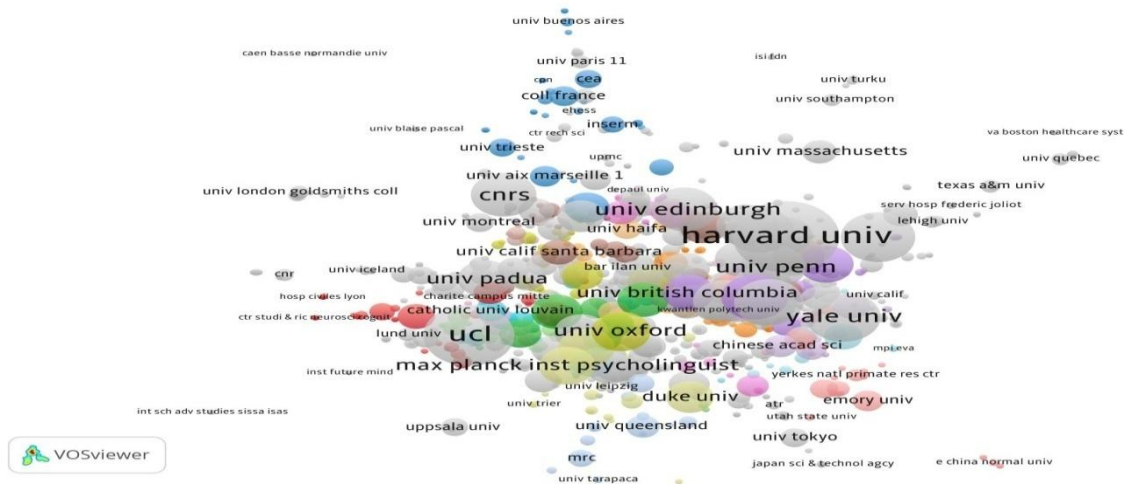

**Figure2.** *Collaboration of Institutions and their clusters*

**Impact and Relevance of Cognition Journal in the Field of Cognitive Science: An Evaluation**

## Analysis of the Publication Output of Top 20 Countries

Table 5 and Figure 3 displays the publication output of the top twenty countries based on number of research publications. After analysing the data it was found that USA acquired 1st rank among the top twenty countries under consideration with its total global citation score 63151. Among all the 61 countries that participated in research during 1999 and 2018, the countries that rank between 2nd and 20th position are: UK, Germany, Canada, France, Italy, Netherlands, Australia, Belgium, Israel, Spain China, Japan, Switzerland, Sweden, Austria, Hungary, South Korea, Finland and New Zealand . By using Country Mapping Analysis, it has been found that the nodes are linked to each other indicating that countries are having collaboration with other associated nations. It could also be identified from the analysis the following countries: USA, UK, Germany, Canada, France, Italy, Netherlands, Australia, Belgium and Israel were identified the most productive countries in terms of the number of research papers published.

**Table5.** *Distribution of the Publication Output of Top 20 Countries*

| S.No. | Country | Records | % | TLCS | TGCS |
|---|---|---|---|---|---|
| 1 | USA | 1390 | 48.43 | 2974 | 63151 |
| 2 | UK | 649 | 22.61 | 1029 | 26264 |
| 3 | Germany | 261 | 9.09 | 289 | 7782 |
| 4 | Canada | 243 | 8.47 | 373 | 9367 |
| 5 | France | 187 | 6.52 | 317 | 8487 |
| 6 | Italy | 133 | 4.63 | 168 | 4534 |
| 7 | Netherlands | 133 | 4.63 | 134 | 4842 |
| 8 | Australia | 105 | 3.66 | 129 | 4304 |
| 9 | Belgium | 87 | 3.03 | 110 | 3957 |
| 10 | Israel | 77 | 2.68 | 62 | 1780 |
| 11 | Spain | 68 | 2.37 | 87 | 2834 |
| 12 | China | 67 | 2.33 | 58 | 1298 |
| 13 | Japan | 61 | 2.13 | 83 | 1941 |
| 14 | Switzerland | 53 | 1.85 | 58 | 1327 |
| 15 | Sweden | 32 | 1.11 | 34 | 720 |
| 16 | Austria | 18 | 0.63 | 14 | 847 |
| 17 | Hungary | 17 | 0.59 | 45 | 564 |
| 18 | South Korea | 17 | 0.59 | 15 | 347 |
| 19 | Finland | 16 | 0.56 | 18 | 523 |
| 20 | New Zealand | 16 | 0.56 | 12 | 265 |

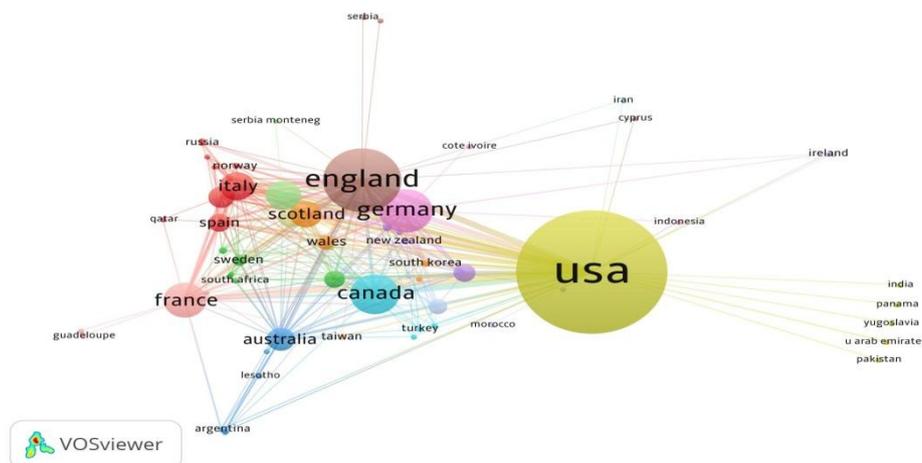

**Figure3.** *Ranking of Country wise Distribution*

## CONCLUSION

The number of papers published in Journal of *Cognition* has gradually increased during 1999–2018 and the study has shown that a total number of 2870 research documents have been published during period of 20 years. The data from this paper also suggest that authors Spelke ES, Carey S, Bloom P, Tanenhaus MK, Tomasello M and Tenenbaum JB were identified as the most prolific authors based on the number of research papers contributed. It could be seen from Institution Wise Analysis that the following institutions: Harvard University, University College London, MIT, Yale University and University of Illinois have published maximum number of research papers in the journal. The following countries: USA, UK, Germany, Canada, France, Italy, Netherlands, Australia, Belgium and Israel were recognised the nations that have contributed highest number of publications during the period under study.